\documentclass{aa}
\usepackage{graphicx}

\def\jh{\mbox{$\rm (J-H)$}}
\def\hk{\mbox{$\rm (H-K)$}}
\def\mM{\mbox{$\rm (m-M)_0$}}
\def\ebv{\mbox{$\rm E(B-V)$}}
\def\ejh{\mbox{$\rm E(J-H)$}}
\def\ebvf{\mbox{$\rm E(B-V)_{FIR}$}}
\def\rc{\mbox{$\rm R_{core}$}}
\def\rcmd{\mbox{$\rm R_{CMD}$}}
\def\rlim{\mbox{$\rm R_{lim}$}}
\def\ms{\mbox{$\rm M_\odot$}}
\def\ds{\mbox{$\rm d_\odot$}}
\def\mj{\mbox{$\rm M_J$}}
\def\jj{\mbox{$\rm J$}}
\def\hh{\mbox{$\rm H$}}

\begin{document}

\title{Discovery of three optical open clusters in the Galaxy}

\author{E. Bica \inst{1}, C. Bonatto \inst{1} \and C. M. Dutra\inst{2}}

\offprints{Ch. Bonatto - charles@if.ufrgs.br}

\institute{Universidade Federal do Rio Grande do Sul, Instituto de F\'\i sica, 
CP\,15051, Porto Alegre 91501-970, RS, Brazil\\
\mail{}
\and
Universidade Estadual do Rio Grande do Sul, Unidade S\~ao Borja, Rua Bompland 
512, S\~ao Borja 97670-000, RS, Brazil\\
}

\date{Received --; accepted --}

\abstract{We report the discovery of three optical open clusters in the Milky Way. 
Two clusters are in Scutum (Cluster\,1 at $\ell=18.44^{\circ}$ and $b=-0.42^\circ$, 
and Cluster\,2 at $\ell=19.60^{\circ}$ and $b=-1.02^\circ$), thus projected 
not far from the Galactic center direction, and the other is in Canis Major 
(Cluster\,3 at $\ell=235.61^{\circ}$ and $b=-4.10^\circ$), near the anti-center direction. 
Cluster\,3 is less populous than Clusters\,1 and 2, but presents evidence of 
being a physical system. The objects were found optically by inspecting maps 
obtained from the Guide Star Catalogue and images from the Digitized Sky Survey. 
No previous identification of cluster has been reported in each area so far. The 
analysis was carried out with 2MASS photometry in J and H. For Cluster\,1 we 
derive an age of $t=25\pm5$\,Myr, a reddening $\ebv=2.18\pm0.03$ and a distance 
from the Sun $\ds=1.64\pm0.19$\,kpc; for Cluster\,2, $t=500\pm100$\,Myr, 
$\ebv=0.91\pm0.03$ and $\ds=2.19\pm0.21$\,kpc; finally for Cluster\,3, 
$t=32-100$\,Myr, $\ebv=0.94\pm0.03$ and $\ds=3.93\pm0.35$\,kpc. Luminosity 
and mass functions are derived for Clusters\,1 and 2 which, in turn, allowed us
to estimate their observed masses as $147\,\ms$ and $89\,\ms$, respectively. 
Estimated total masses, by extrapolating the mass functions to $0.08\,\ms$, 
amount to $382\,\ms$ and $614\,\ms$, for the two clusters. Cluster\,3 has an 
observed mass of $\sim55\,\ms$. The present results indicate that further 
searches in the optical might still reveal new open clusters, and more so 
in infrared bands. 

 }

\titlerunning{New open clusters in the Galaxy}

\authorrunning{E. Bica et al.}

\maketitle

\section{Introduction}

The discovery of new physical stellar systems in the Galaxy is essential for
the task of completing their census. This, along with the determination of 
properties such as distance, age, mass and dynamical state, and spatial
distribution and evolutionary processes in the disk (e.g. Janes \& Adler
\cite{JA1982}, Twarog et al. \cite{Twa1997}) will give a better 
statistical insight on their formation, evolution and eventual disruption. 
Open clusters span a wide range of ages from the very young to the intermediate 
age ones, whose ages and distances can be determined by fitting isochrones to 
their colour-magnitude diagrams (CMDs), with a precision depending on the depth 
of the photometry and field contamination. These objects are formed along the 
Galactic plane where there is an abundance of gas and dust and their orbits 
become perturbed by the cumulative effect of passages near interstellar clouds. 
Due to their spatial location, young open clusters 
can be considered as tracers of the Galaxy's spiral structure (Chen et al. 
\cite{Chen2003}). Thus, the discovery and characterization of additional open 
clusters represent a step further in the understanding of Galaxy structure and 
Galaxy formation processes.

In the recent decades some new open clusters have been discovered in the
optical domain, e.g. Pfleiderer et al. (\cite{Pfleiderer1977}), Turner et al. 
(\cite{Turner1986}) and Saurer et al. (\cite{Saurer1994}). Recently, two 
optical open clusters in Cygnus OB2 have been found and their properties 
determined (Bica et al. \cite{BBD2003}). In the infrared, the amount of 
new embedded clusters is striking, mostly due to the recent release of the 
Two Micron All Sky Survey (hereafter 2MASS, Skrutskie et al. \cite{2mass1997}) 
catalogue and atlas. We point out the recent discovery of 346 embedded clusters 
and candidates along the spiral arms of the Galaxy (Dutra et al. \cite{Dutra2003a}, 
Bica et al. \cite{Bica2003}).

In the present study we report the finding of three new optical open clusters.
Their properties will be derived by means of 2MASS photometry, since these
optical objects show up clearly in the near-infrared. In addition, 
the uniform and essentially complete sky coverage provided by 2MASS allows one 
to properly take into account background regions with suitable star count 
statistics, which is fundamental in order to correctly identify and characterize 
the stellar content of clusters.

In Section~\ref{The3NOC} we present the new open clusters and show the optical images 
where the clusters have been found. In Sect.~\ref{2massPh} we obtain the 2MASS 
photometry and introduce the $\jj\times\jh$ CMDs. In Sect.~\ref{StructAnal} we discuss 
the radial density distribution of stars and derive structural parameters for the 
clusters. In Sect.~\ref{Fund_par} we fit isochrones to the near-infrared CMDs and 
derive cluster parameters. In Sect.~\ref{LumFunc} we derive the luminosity and mass 
functions (hereafter LF and MF) and estimate the stellar masses of each cluster.  
Concluding remarks are given in Sect.~\ref{Conclu}.

\section{The three new open clusters}
\label{The3NOC}

The clusters have been found during systematic inspections of the Milky Way 
with maps generated by means of the Guide Star Catalogue, Sky Survey
Charts and detailed charts of candidate regions with Digitized Sky
Survey (DSS and XDSS) fields. Equatorial and Galactic coordinates, and
angular diameters are given in Table~\ref{tab1}. 

Clusters\,1 and 2 are in Scutum, thus projected not far from the direction
of the Galactic center, while Cluster\,3 is in Canis Major, not too far 
from the anti-center. None of these objects is listed in previous catalogues
(Alter et al. \cite{Alter1970}; Lyng\aa~\cite{Lyngaa1987}; Dias et al. 
\cite{Dias2002}).

\begin{table*}
\caption[]{Observational parameters for the new clusters.}
\label{tab1}
\renewcommand{\tabcolsep}{4.6mm}
\begin{tabular}{cccccccc}
\hline\hline
Object&$\alpha(2000)$&$\delta(2000)$&$\ell$&$b$&Diameter&\rcmd&\ebvf\\
\hline
Cluster\,1& 18$^h$26$^m$04$^s$ & $-13^\circ$03\arcmin32\arcsec&$~18.44^\circ$&$-0.42^\circ$&$3.5\arcmin\times2.5\arcmin$&1.5\arcmin&15.77\\
Cluster\,2& 18$^h$30$^m$30$^s$ & $-12^\circ$18\arcmin59\arcsec&$~19.60^\circ$&$-1.02^\circ$&$6.0\arcmin\times6.0\arcmin$&2.0\arcmin&~5.63\\
Cluster\,3& 07$^h$19$^m$07$^s$ & $-22^\circ$01\arcmin40\arcsec&$235.61^\circ$&$-4.10^\circ$&$2.5\arcmin\times2.5\arcmin$&1.1\arcmin&~1.66\\
\hline
\end{tabular}
\begin{list}{Table Notes.}
\item Column~6 gives the optical diameters; Column~7 gives the extraction radius used
in the CMD analyses; \ebvf\ is Schlegel et al.'s (\cite{Schlegel1998}) reddening values,
derived from the far-infrared dust emission.
\end{list}
\end{table*}

\begin{figure} 
\resizebox{\hsize}{!}{\includegraphics{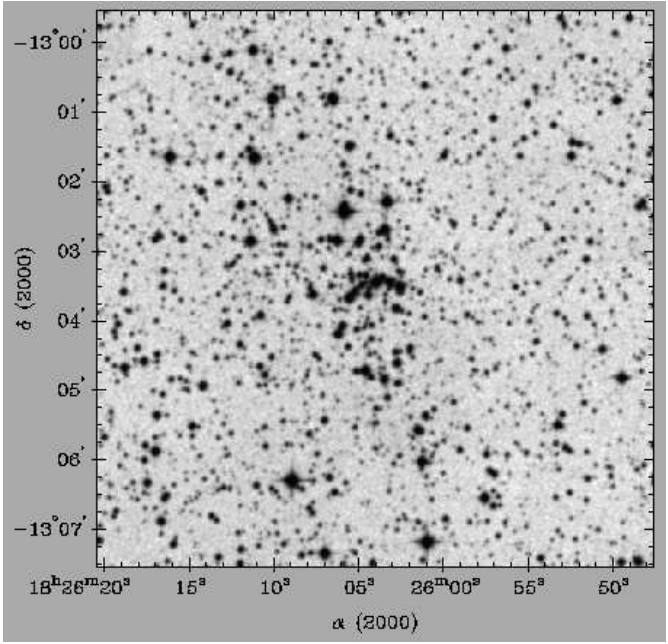}}
\caption[]{8$^{\prime}$ $\times$ 8$^{\prime}$ XDSS R image of Cluster\,1.}
\label{fig1}
\end{figure}

\begin{figure} 
\resizebox{\hsize}{!}{\includegraphics{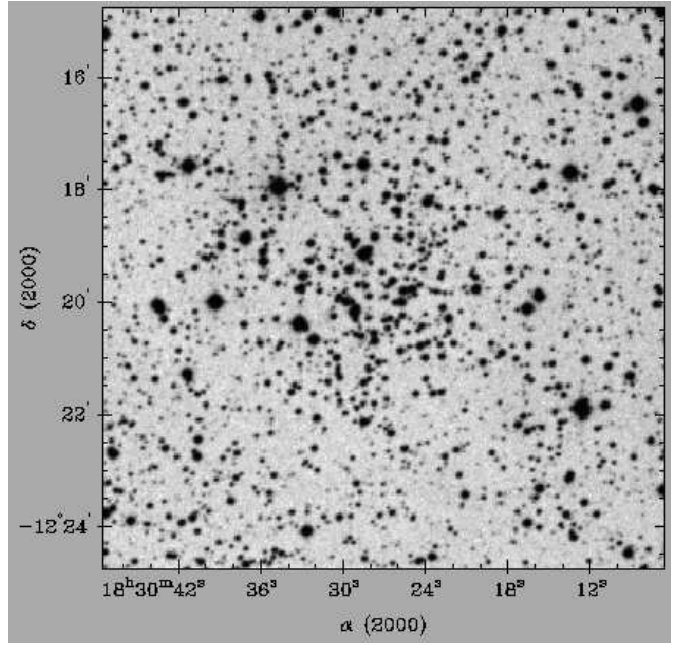}}
\caption[]{10$^{\prime}$ $\times$ 10$^{\prime}$ DSS B image of Cluster\,2.}
\label{fig2}
\end{figure}

\begin{figure} 
\resizebox{\hsize}{!}{\includegraphics{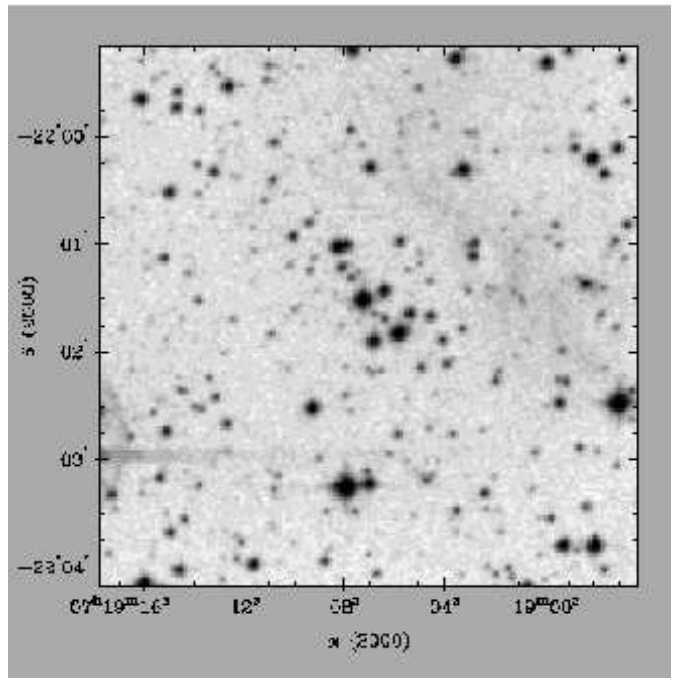}}
\caption[]{5$^{\prime}$ $\times$ 5$^{\prime}$ XDSS R image of Cluster\,3.}
\label{fig3}
\end{figure}

We show in Figs.~\ref{fig1}, \ref{fig2} and \ref{fig3} the three open clusters 
in optical bands. Clusters\,1 and 2 are projected against rich fields, as
expected from their Galactic coordinates.
Considering the stellar distributions, the three objects  stand out from 
the background areas. Open Cluster\,1 is elongated in the North-South direction 
and rather concentrated to the center (Fig.~\ref{fig1}). Although Cluster\,2 is 
looser than Cluster\,1 it still detaches from the rich background (Fig.~\ref{fig2}). 
Finally, Cluster\,3, although poorer in stars, presents a central concentration. 

Cluster\,1 is projected approximately 16\arcmin\ to the North-East of the \ion{H}{ii}
region Sh\,2-53 (Sharpless \cite{Sharp1959}). Cluster\,2 lies at approximately 
13\arcmin\ West of the open cluster Ruprecht\,141 (Alter et al. \cite{Alter1970}) 
which has no available parameters, according to the open cluster database WEBDA 
(Mermilliod \cite{Merm1996} --- {\em http://obswww.unige.ch/webda}). Finally, Cluster\,3 
is projected approximately 17\arcmin\ Southeast of the young open cluster NGC\,2367
--- with age $\approx5$\,Myr (Vogt \& Moffat \cite{Vogt1972}) --- and 5\arcmin\ South
of the \ion{H}{ii} region RCW\,14 (Rodgers et al. \cite{Rodgers1960}). RCW\,14 appears 
to be the nucleus of the larger nebula Brand\,16 (Brand et al. \cite{BBW1986}).
The projected environments of Clusters\,1 and 3 thus contain young objects.

\section{The 2MASS photometry}
\label{2massPh}

\begin{figure} 
\resizebox{\hsize}{!}{\includegraphics{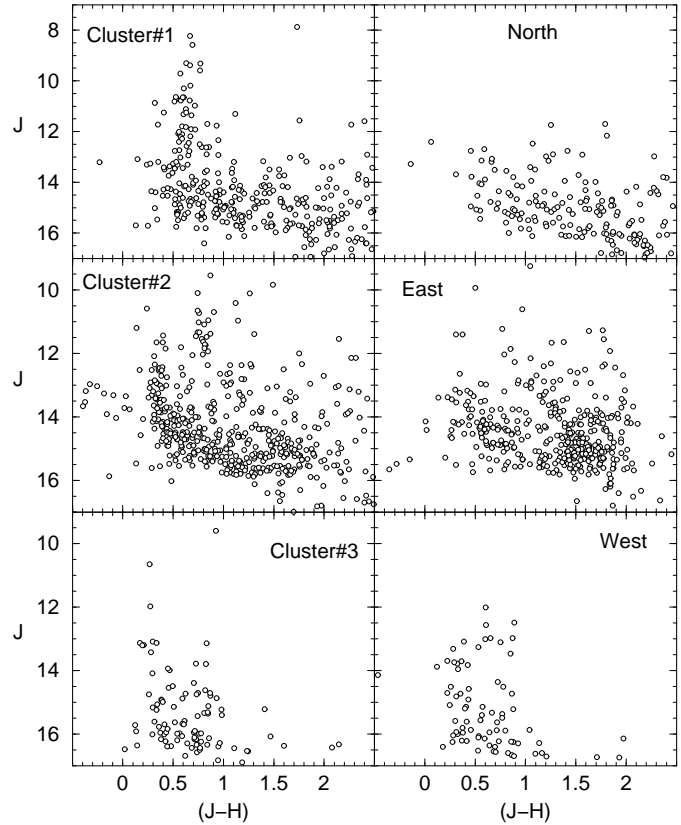}}
\caption[]{$\jj\times\jh$ CMDs for the new clusters (left panels); representative 
offset fields are also shown (right panels).}
\label{figCMD}
\end{figure}

J and H photometry has been obtained from the 2MASS All Sky data release,
available at {\em http://www.ipac.caltech.edu/2mass/releases/allsky/}. 2MASS
photometric errors typically attain 0.10\,mag at $\jj\approx16.2$ and
$\hh\approx15.0$, see e.g. Soares \& Bica (\cite{SB2002}). For 
each cluster we made circular extractions centered on the coordinates given
in Table~\ref{tab1} with the extraction radius (\rcmd) listed in Column~7. 
We decided for an extraction area smaller than the optical one 
in order to minimize background contamination and increase the membership probability 
of the stars sampled. Comparison fields have been extracted inside 
circular areas with the same radii as those used for the clusters in four positions 
at North, South, East and West of each cluster, with center to center distances 
corresponding to $3\times\rcmd$. Extractions have been performed using the VizieR 
tool at {\em http://vizier.u-strasbg.fr/viz-bin/VizieR?-source=2MASS}. In 
Fig.~\ref{figCMD} we show the $\jj\times\jh$ CMDs for each cluster (left panels) 
along with a representative offset field (right panels).

In the fields of Clusters\,1 and 2 star colours reach as
much as $\jh=3$, beyond Fig.~\ref{figCMD} limits, while for the Cluster\,3 
area all stars are included in the figure. These very red colours, which 
correspond to high reddening values, are caused by accumulation of dust in 
the line of sight, since we are dealing with very low Galactic latitudes for 
Clusters\,1 and 2 (Table~\ref{tab1}). This is confirmed by Schlegel et al.'s 
(\cite{Schlegel1998}) reddening values derived from dust emission in the 
far-infrared (FIR), given in the last column of Table~\ref{tab1}. Dutra \& Bica 
(2002) studied low-latitude directions using star clusters as probes for reddening 
in the foreground and background of the clusters. More recently, Dutra et al. 
(\cite{Dutra2003b}) found evidence of the need of a calibration correction by 
a multiplicative factor of 0.75 to Schlegel et al.'s values, at least for low 
Galactic latitudes towards the Galactic center (down to $|b|=4^{\circ}$). Anyhow, 
the \ebvf\ values in the directions of Clusters\,1 and 2 would still be very high.

Cluster\,1 presents a prominent main sequence (MS) as compared to the corresponding
offset field (Fig.~\ref{figCMD}), which is indicative of a young age. Cluster\,2 exhibits 
a MS and giants, indicating older ages. Cluster\,3 presents evidence of
a young MS. 

\section{Cluster structure}
\label{StructAnal}

\begin{figure} 
\resizebox{\hsize}{!}{\includegraphics{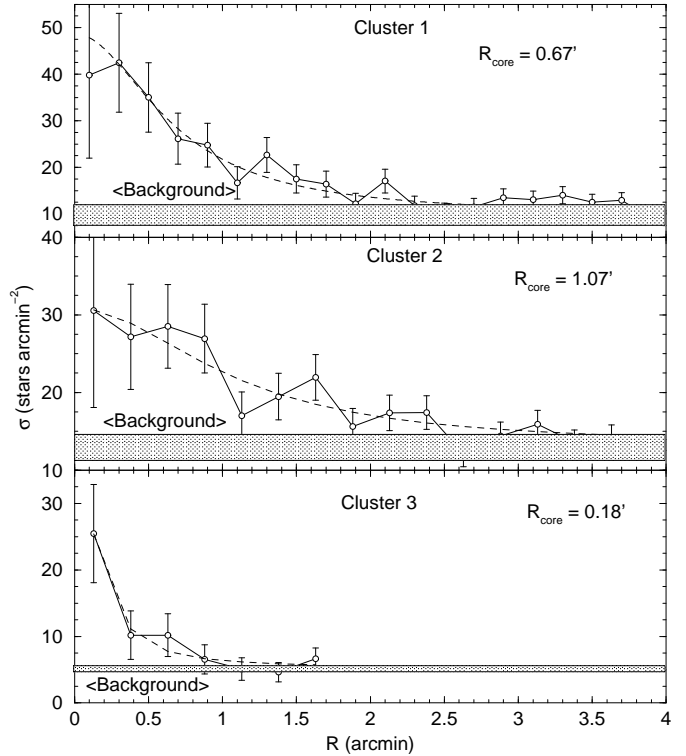}}
\caption[]{Radial distribution of surface star density. The average background levels
are shown as shaded rectangles; Poissonic errors are also shown. For Clusters\,1
and 2, magnitude ($\jj<15.5$) and colour cutoffs have been applied to the object and 
offset fields. The dashed lines show a surface density profile fit to the radial
distribution of stars; the resulting core radius for each cluster is indicated.}
\label{figRAD}
\end{figure}

For the specific purpose of better accessing the overall cluster structure, 
we made additional star extractions reaching as far as the optical diameter of each 
cluster, according to the values in Table~\ref{tab1}. With the stars obtained in these 
new extraction areas, we built the star density radial distributions, defined as the
number of stars per area, in and around the clusters, which are shown in 
Fig.~\ref{figRAD}. 

Before counting stars, we applied a cutoff ($\jj<15.5$) to Clusters\,1 and 2
and their corresponding offset fields to avoid undersampling, i.e. to avoid spatial 
variations in the number of faint stars which are numerous, affected by large errors, 
and may include spurious detections, in the area of the clusters. Colour filters have 
also been applied to both Clusters\,1 and 2 and offset fields, in order to account 
for the contamination of the Galaxy -- only stars with colour in the range
$0.0\leq\jh\leq1.1$\ have been considered. This procedure has been applied in the 
analysis of the open cluster M\,67 (Bonatto \& Bica \cite{BB2003}). Due to the relatively 
small number of stars in the area of Cluster\,3, no cutoff has been applied. The radial 
distribution has been determined by counting stars inside concentric annuli with a step 
of 0.25\arcmin\ in radius up to the new extraction limits. The background contribution,
shown in Fig.~\ref{figRAD} as shaded rectangles, corresponds to the average number of 
stars included in the four offset fields.

Cluster\,1 has a slight deficiency of stars near the center as compared to the
neighbouring annulus, but beyond R$=0.25\arcmin$ it presents a well-defined and 
rather smooth profile (top panel) with star counts well above the background, 
considering the Poissonic errors. The central deficiency might in part be 
accounted for by faint star images blended to the several bright ones near the 
object center. According to the radial distribution of stars, Cluster\,1 extends
beyond the CMD extraction radius, reaching a limiting radius of $\rlim\approx2.7\arcmin$. 
The same is true for Cluster\,2 (middle panel), although its profile is not as smooth 
as that of Cluster\,1. Cluster\,2 extends to $\rlim\approx3.5\arcmin$. The 
well-defined and smooth profile of Cluster\,3 (bottom panel) is more concentrated 
than both Clusters\,1 and 2, extending to $\rlim\approx1.6\arcmin$.

Although the clusters' spatial geometry may not be perfectly spherical, we still can 
apply King's law (\cite{King1966}) in order to derive first order structural parameters. 
A cluster core radius \rc\ can be calculated by fitting a King's surface density 
profile $\sigma(R)=\frac{\sigma_0}{1+\left(R/\rc\right)^2}$\ to the background-subtracted 
radial distribution of stars. The resulting fits are also shown in Fig.~\ref{figRAD}, 
as dashed lines. Clusters\,1 and 3 follow within uncertainties a King profile, with 
$\rc=0.67\arcmin$\ and $\rc=0.18\arcmin$, respectively. The radial
density profile of Cluster\,2 is not as smooth as those of Clusters\,1 and 3,
consequently, the King's fit is not as good and the resulting $\rc\sim1.07\arcmin$\
has to be taken as an estimate only. Using the cluster distances derived in 
Sect.~\ref{Fund_par} below, the linear core radii turn out to be $\rc=0.32\pm0.03$\,pc, 
$\sim0.68\pm0.06$\,pc and $0.21\pm0.02$\,pc, respectively for Clusters\,1, 2 and 3.
Finally, the angular diameters of 5.4\arcmin, 7.0\arcmin\ and 3.2\arcmin\
(Table~\ref{tab1}), convert to linear limiting diameters of $2.6\pm0.3$\,pc, 
$4.4\pm0.4$\,pc and $3.8\pm0.3$\,pc. Cluster\,1 is very young (Sect.~\ref{Fund_par}), 
indeed its diameter is comparable to typical values observed in infrared embedded 
clusters (Bica et al. \cite{Bica2003}). The older Clusters\,3 and especially 2, have  
larger diameters which must be reflecting the stochastic effects of the Galactic tidal 
processes (Bonatto \& Bica \cite{BB2003}). In the case of the much older open cluster 
M\,67, located about 1\,kpc outside the Solar circle, the limiting diameter is 
$\approx12$\,pc, while Cluster\,2 with a limiting diameter of $\approx4.4$\,pc,
is located $\approx2$\,kpc inside the Solar circle.

\section{Fundamental parameters}
\label{Fund_par}

In the following two sections we will base our analyses on stars extracted within 
\rcmd\ (Table~\ref{tab1}). In order to derive cluster parameters we use solar 
metallicity Padova isochrones from Girardi et al. (\cite{Girardi2002}) computed with 
the 2MASS J, H and K$_S$ filters (available at 
{\em http://pleiadi.pd.astro.it/$\sim$lgirardi/$-$isoc$\_$photsys.00/}). The 2MASS 
transmission filters produced isochrones very similar to the Johnson ones, with 
differences of at most 0.01 in \jh\ (Bonatto et al. \cite{BBG2004}). For reddening 
and absorption transformations we use R$_V$ = 3.2, and the relations A$_J = 0.276$A$_V$ 
and $\ejh=0.33\ebv$, according to Dutra et al. (\cite{DSB2002}) and references
therein. 

\begin{figure} 
\resizebox{\hsize}{!}{\includegraphics{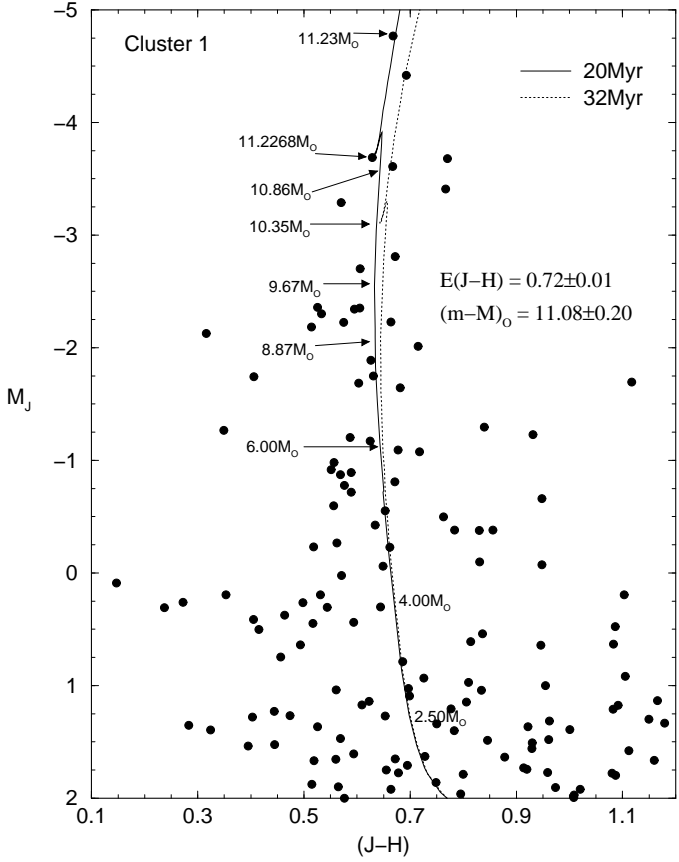}}
\caption[]{Isochrone fitting to the $\mj\times\jh$ CMD of Cluster\,1. Equally good 
fits are obtained for 20 and 32\,Myr isochrones, resulting in $\ebv=2.18\pm0.03$ 
and $\ds=1.64\pm0.19$\,kpc. Stellar masses along the main-sequence are indicated.}
\label{figIsocC1}
\end{figure}

We show in Fig.~\ref{figIsocC1} the isochrone fitting to the $\mj\times\jh$ CMD
of Cluster\,1. The \mj\ values are obtained after applying the distance modulus
derived below for each cluster. The observed scatter of stars in colour along the
MS makes it possible to fit it equally well with the 20 and 32\,Myr isochrones, 
thus constraining the age to this  narrow range. The fit and related 
uncertainties give a distance modulus $\mM=11.08\pm0.20$, 
$\ebv=2.18\pm0.03$ and $\ds=1.64\pm0.19$\,kpc. 
Representative stellar masses are indicated along the 20\,Myr isochrone.

\begin{figure} 
\resizebox{\hsize}{!}{\includegraphics{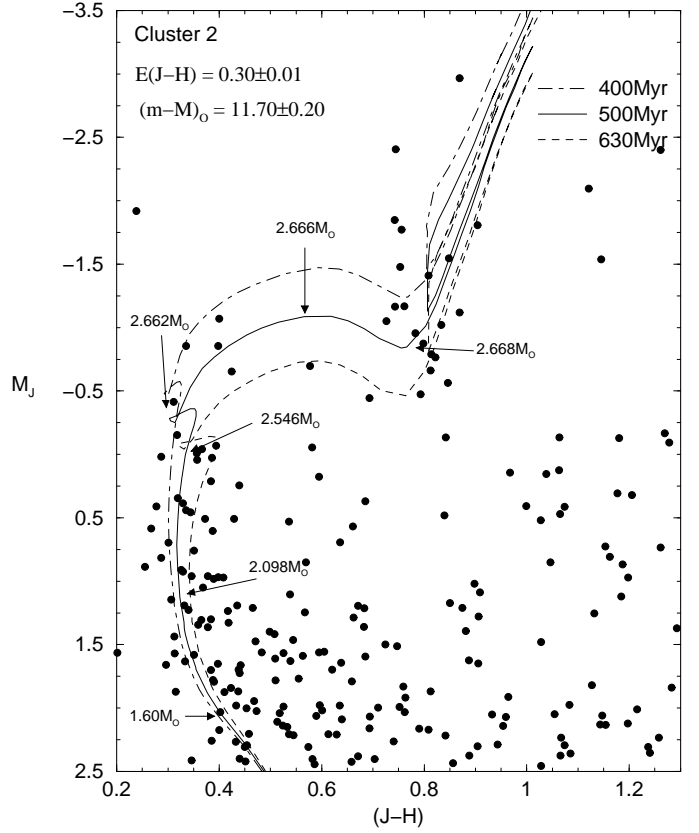}}
\caption[]{Same as Fig.~\ref{figIsocC1} for Cluster\,2. Acceptable fits are obtained
for 400--630\,Myr isochrones, resulting in $\ebv=0.91\pm0.03$ and 
$\ds=2.19\pm0.21$\,kpc.}
\label{figIsocC2}
\end{figure}

The presence of giants in the CMD of Cluster\,2 indicates an older age for this 
object. Indeed, allowing for the star scatter, the distribution of stars can be 
fitted by isochrones with ages in the range 400 -- 630\,Myr, as can be 
seen in Fig.~\ref{figIsocC2}. For Cluster\,2 we derive $\mM=11.70\pm0.20$, 
$\ebv=0.91\pm0.03$ and $\ds=2.19\pm0.21$\,kpc.

\begin{figure} 
\resizebox{\hsize}{!}{\includegraphics{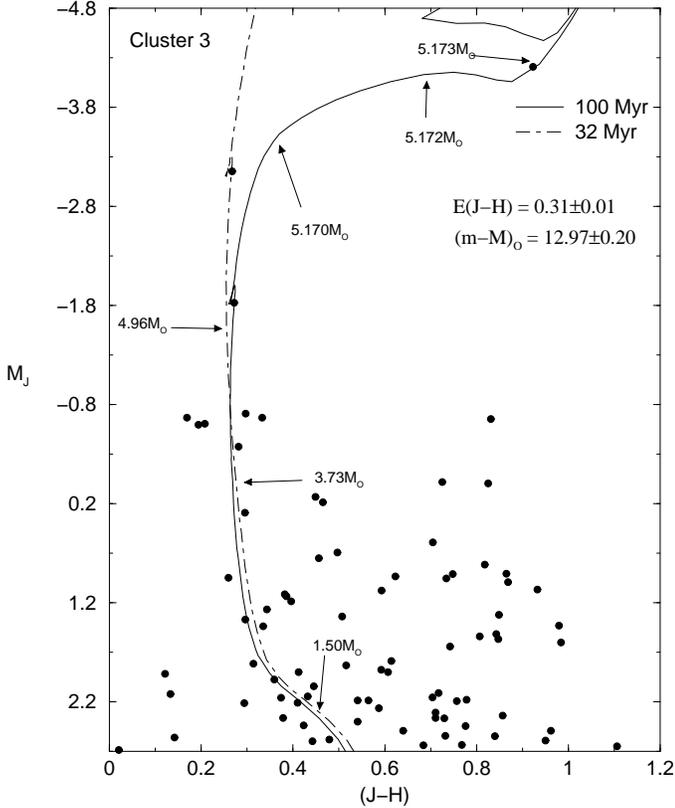}}
\caption[]{Same as Fig.~\ref{figIsocC1} for Cluster\,3. Depending on the
membership of a single evolved star, acceptable fits are obtained
for 100 and 32\,Myr isochrones, resulting in $\ebv=0.94\pm0.03$ and 
$\ds=3.93\pm0.35$\,kpc.}
\label{figIsocC3}
\end{figure}

The best-fit for Cluster\,3 is obtained with the 100\,Myr isochrone, however,
this solution is strongly based on the membership assumption of a single evolved star.
Arguing in favour of membership, the coordinates of that bright star -- $\alpha(2000) = 
07^h01^m01.10^s$\ and $\delta(2000) = -22^\circ01\arcmin29.57\arcsec$ -- place it at 
0.18\arcmin\ from the cluster's central position. Another solution, disconsidering 
the bright star, is obtained with the 32\,Myr isochrone. Both solutions are shown 
in Fig.~\ref{figIsocC3}. In any case, proper-motion information is necessary to 
derive the precise age for this cluster. For Cluster\,3 we derive $\mM=12.97\pm0.20$,
$\ebv=0.94\pm0.03$ and $\ds=3.93\pm0.35$\,kpc.

The \ebvf\ reddening values in Table~\ref{tab1} are overestimations,
since they represent the dust column contribution integrated along the whole line-of-sight 
(up to the disk edge). Thus, using the exponential dust distribution model of Chen et al. 
(\cite{Chen1999}) and the distances derived from the CMDs, we obtain the following
foreground reddening values for Clusters\,1, 2 and 3, respectively: $\ebv=0.92$, 
1.18 and 1.17. The results for Clusters\,2 and 3 are close to the reddening values derived 
from the CMDs, suggesting that for the directions to these clusters an exponential dust
distribution law is a good approximation. Cluster\,1 lies at a very low latitude and as
expected the agreement between reddening estimates is not good.

Cluster\,1, at 1.6\,kpc, does not appear to be related to the \ion{H}{ii}
region complex Sh2-53, since this object is estimated to be at a kinematic
distance of $\approx4.3$\,kpc (Georgelin \& Georgelin \cite{GG1976}). Cluster\,1 
seems to be related to the Sgr-Car arm, while Sh2-53, to the Sct-Cru arm. 
Cluster\,3, at 3.9\,kpc, appears to be located in the background of the closely
projected young cluster  which is at an estimated distance of 2\,kpc (Vogt \& 
Moffat \cite{Vogt1972}), as well as in the background of the \ion{H}{ii} region 
Brand\,16 or RCW\,14, located at 1.9\,kpc from the Sun (Brand \& Blitz \cite{BrBl1993}).
The latter nebulae are probably related to the opne cluster NGC\,2367.

The fact that Cluster\,3 is projected behind the nebula Brand\,16 poses a question
whether the cluster might be very young and physically related to it and RCW\,14.
To address this question we plotted in a $\jh\times\hk$ diagram  the stars inside
\rcmd\ and the corresponding offset field extractions. This colour-colour
diagram is sensitive to reddening and the presence of excess emission in the
K band, related to protostellar disks (Soares \& Bica \cite{SB2002} and 
Soares \& Bica \cite{SB2003}, and references therein). We applied a cutoff of
$\jj<15.5$ in order to avoid uncertainties larger than 0.08 in \jj. This
procedure avoids as well prohibitive errors in colours. As a result, we found
no \hk\ excess stars associated to Cluster\,3. This result suggests that
Cluster\,3 is indeed in the background of Brand\,16.

\section{Luminosity and mass functions}
\label{LumFunc}

Observed star counts as a function of magnitude (or mass) can be compared to theoretical 
predictions of number density as a function of luminosity (or mass). This can be 
used to test whether stars are present in numbers as expected and to estimate the 
cluster's total mass, taking into account stellar masses as low as $0.08\,\ms$. 
This relation between number density and luminosity (or mass) is referred to as the 
LF (or MF) -- Salpeter (\cite{Sal1955}).

The accurate determination of a cluster's LF (or MF) suffers from some problems, in 
particular {\it (i)} the contamination of cluster members by field stars, {\it (ii)} 
the observed incompleteness at low-luminosity (or low-mass) stars, and  {\it (iii)} 
the mass-segregation, which may affect even poorly populated, relatively young
clusters (Scalo \cite{Scalo1998}). The 2MASS uniform sky coverage allows one to overcome, 
at least in part, points {\it (i)} -- since suitable offset fields can be selected 
around the cluster and {\it (iii)} -- the entire cluster area can be included in 
the analyses. Thus, advanced stages of mass-segregation would affect more significantly
the analysis of very old, dynamically evolved clusters (e.g. M\,67, Bonatto \& Bica 
\cite{BB2003}). This is not the case of the three new clusters dealt with in the
present work.

\begin{figure}
\resizebox{\hsize}{!}{\includegraphics{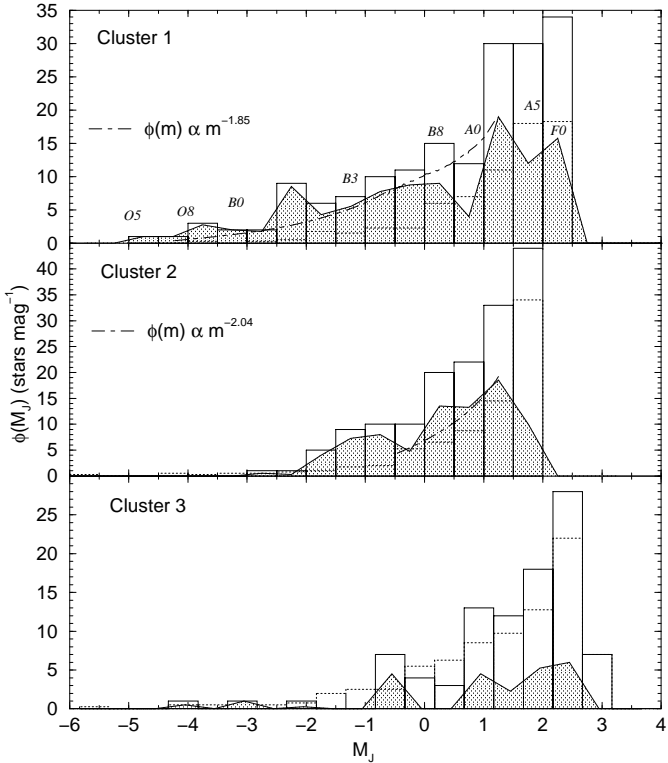}}
\caption[]{Luminosity functions ($\phi(\mj)$) in terms of the absolute magnitude \mj. 
For Clusters\,1 and 2, magnitude ($\jj<15.5$) and colour cutoffs have been applied to
the object and offset fields. Continuous line: star counts in the cluster's area;
dotted line: star counts in the offset fields; shaded area: background-subtracted
LF. Representative MS spectral types, taken 
from Binney \& Merrifield (\cite{Binney1998}), are indicated in the top panel.
MF fits ($\phi(m)\propto m^{-(1+\chi)}$) are shown as dot-dashed lines.}
\label{figLF}
\end{figure}

Figure~\ref{figLF} depicts the LFs ($\phi(\mj)$) in the J filter (shaded area) for the 
three new clusters, built as the difference of the number of stars in a given magnitude 
bin between object (continuous line) and average offset field (dotted line). The LFs 
are given in terms of the absolute magnitude \mj, after applying the distance modulus 
derived in Sect.~\ref{Fund_par} for each cluster.

We remind that the LFs for Clusters\,1 and 2 (Fig.~\ref{figLF}) are built after applying 
magnitude ($\jj<15.5$) and colour cutoffs to the objects and offset fields 
(Sect.~\ref{Fund_par}).

The background-subtracted LFs of Clusters\,1 and 2 present significant 
star excesses over the background, increasing up to $\jj\approx14.2$\ for Cluster\,1, and 
$\jj\approx13.2$\ for Cluster\,2. Due to the different distances of both
clusters, the turnover in their LFs begins at $\mj\approx1.25$, 
corresponding to the spectral types A\,2--A\,3. It should be noted that the bump at 
$-2.0\leq\mj\leq -0.5$ in the LF of Cluster\,2 can be accounted for by 
the overlap in \mj\ of MS stars near the turnoff at $0.3\leq\jh\leq 0.4$ with 
red giant stars at $0.7\leq\jh\leq 0.9$ (Fig.~\ref{figIsocC2}).
Despite the scarcity of stars in Cluster\,3, this object still presents 
star excesses over the background which roughly increase towards the low-mass end 
(bottom panel), peaking at $\jj\approx16.2$\ ($\mj\approx2.4$, F\,0). Spectral 
types and corresponding \mj\ for MS stars, taken from Binney \& Merrifield 
(\cite{Binney1998}), are displayed on the top panel of Fig.~\ref{figLF}.

An estimate of the mass presently stored in stars in a cluster can be obtained by 
fitting the function $\phi(m)=\frac{dN}{dm}\propto m^{-(1+\chi)}$ to the cluster's
MF, restricted to the MS and turnoff. Then, the number of member stars is
$N^*=\int_{m_{low}}^{m_{high}} \phi(m)dm$, and the stellar mass is
$M^*=\int_{m_{low}}^{m_{high}} m \phi(m)dm$, where $m_{low}$ and $m_{high}$ define 
the range of stellar masses from the turnover at the low-MS up to the turnoff. The 
stellar mass-luminosity correspondence, necessary to convert the LFs 
in Fig.~\ref{figLF} into MFs, $\phi(m)=\phi(\mj)|\frac{dm}{d\mj}|^{-1}$, 
has been obtained from the Padova isochrones which best fit the CMDs (Sect.~\ref{Fund_par}): 
20\,Myr for Cluster\,1, 500\,Myr for Cluster\,2 and 100\,Myr for Cluster\,3.

\begin{table*}
\caption[]{Mass-function fit and related parameters.}
\label{tabLFM}
\renewcommand{\tabcolsep}{2.95mm}
\begin{tabular}{ccccccccccc}
\hline\hline
&\multicolumn{3}{c}{Observed}&&&&&\multicolumn{3}{c}{Calculated}\\
\cline{2-4}\cline{9-11}
Object&N$^*_{MS}$&N$^*_{evol}$&M$^*_{evol}$&Fit&$\chi$&$m_{low}$&$m_{high}$&M$^*$&N$^*_{max}$&M$^*_{max}$\\
&(stars)&(stars)&(\ms)&(\mj)&$\left(m^{-(1+\chi)}\right)$&(\ms)&(\ms)&(\ms)&(stars)&(\ms)\\
\hline
Cluster\,1 & 103&~0 & ~0&$-4.75\rightarrow 1.25$ & 0.85 & 2.515 & 11.22& 147  & ~755&382\\
Cluster\,2 & ~62&18 & 48&$-0.25\rightarrow 1.25$ & 1.04 & 2.037 & 2.662& ~89$\dag$ & 2036$\dag$&614$\dag$\\
Cluster\,3 & ~25&~1 & $\sim5.2$&      ---      &  --- &   --- &  --- & ~55$\dag$ & ---& ---\\
\hline
\end{tabular}
\begin{list}{Table Notes.}
\item Column~5 is the \mj\ range to which the MF has been fitted. The values for 
Cluster\,3 are derived from directly counting stars in each magnitude bin. ($\dag$): includes
MS and evolved stars. N$^*_{max}$ and M$^*_{max}$ are calculated by extrapolating the
MF fits to $m_{low}=0.08\,\ms$.
\end{list}
\end{table*}

Since Cluster\,1 is young with no late-type stars observed in the CMD, we included in the 
fit the whole range of observed masses down to the turnover at $\mj\approx1.25$, which 
corresponds to a mass m$=2.51\,\ms$. We point out that for high masses we are not dealing
with OV stars (Fig.~\ref{figIsocC1}), in fact these very few stars are hot giants which
do not significantly affect the analysis below. To estimate the mass of Cluster\,2 we took 
into account separately the MS and evolved stars. We first fitted a MF to the MS stars 
from the turnover at $\mj=1.25$ (m$=2.037\,\ms$) up to $\mj=-0.25$ (m$=2.615\,\ms$) 
to avoid contamination by the overlapping giants, and then extrapolated this function to 
the turnoff ($\mj\approx-0.50$, m$\approx2.662\,\ms$). The mass of the giants (M$^*_{evol}$)  
has been estimated by counting the number of stars in each magnitude bin (after 
subtracting the MS stars in the overlap region) and multiplying this value by the average 
mass of the giants included in the bin. M$^*_{evol}$ is given in column~4 of Table~\ref{tabLFM}.
The resulting MF fits for Clusters\,1 and 2, transformed back to $\phi(\mj)$, are shown as 
dot-dashed lines in Fig.~\ref{figLF}. The related parameters are given in Table~\ref{tabLFM}. 
We also include in Table~\ref{tabLFM} the number of observed MS (N$^*_{MS}$) and
evolved (N$^*_{evol}$) member stars, obtained by summing up the stars in each magnitude 
bin in the corresponding LFs. 
 
The MF method underestimates both the mass and number of stars, since 
completeness effects affect the 2MASS observations in the low-mass end. More 
realistic values can be obtained by extrapolating the MFs derived above down to 
the theoretical stellar low-mass end $m_{low}=0.08\,\ms$. The corresponding maximum 
number of stars (N$^*_{max}$) and mass (M$^*_{max}$) for Clusters\,1 and 2 are given 
in the last two columns of Table~\ref{tabLFM}, respectively.

Due to the small number of stars in the CMD, we could not apply the MF fit method to 
Cluster\,3. Thus, we estimate the stellar mass in Cluster\,3 by adopting an average 
mass per magnitude bin (based on the mass-luminosity correspondence taken from the 
100\,Myr Padova isochrone) and multiplying this value by the number of stars in each 
bin. 

The MF slopes derived for Clusters\,1 and 2 (Table~\ref{tabLFM}) are comparable to, 
but somewhat flatter than, a standard Salpeter slope of $\chi=1.35$ (Binney \& 
Merrifield \cite{Binney1998}). It is interesting to note that the extrapolated mass 
of Cluster\,2 turns out to be larger than that of Cluster\,1, in the opposite sense 
with respect to the observed mass. This fact can be accounted for by the steeper MF 
slope ($\chi=1.04$) of Cluster\,2 than that of Cluster\,1 ($\chi=0.85$).

None of the present objects is a massive open cluster (in the sense of say,
M$^*\ge10^3\,\ms$). Cluster\,3 may be an evolutionary product of low-mass
embedded clusters such as those described by Soares \& Bica (\cite{SB2002} and
\cite{SB2003}).

\section{Concluding remarks}
\label{Conclu}

The recent discoveries of star clusters in the Galactic disk have shown that 
the census of these objects is not complete, even in the optical domain. 
In this paper we report the discovery of three optical open clusters in the Galaxy.  
Two clusters are in Scutum, at very low Galactic latitudes, and the other in Canis 
Major. The present photometric and structural analyses make use mostly of J and H 
2MASS All Sky data release photometry.

Arguments in favour of their nature as open clusters are: {\it (i)} the visual 
contrast between the stellar concentrations in the cluster and background regions
(Figs.~\ref{fig1}, \ref{fig2} and \ref{fig3}); {\it (ii)} the CMDs which 
present well-defined sequences and are different from the offset field ones
(Fig.~\ref{figCMD}); {\it (iii)} the radial star density profiles which show 
important excesses over the background and can be fitted by a King's law 
(Fig.~\ref{figRAD}); {\it (iv)} the constrained solutions of the isochrone 
fits to the CMDs, allowing a narrow range of ages (Figs.~\ref{figIsocC1}, 
\ref{figIsocC2} and \ref{figIsocC3}); and {\it (v)} the background-subtracted LFs 
which present significant star excesses over the background down to A\,2--A\,3 
stars for Clusters\,1 and 2, and to F\,0 stars for Cluster\,3 (Fig.~\ref{figLF}). 
Completeness effects and photometric errors do not allow inferences on lower masses. 
For the sampled magnitude range, the number of member stars turns out to be 103, 80 
and 26, respectively for Clusters\,1, 2 and 3. 

Fundamental parameters for the objects have been derived by fitting solar-metallicity
Padova isochrones to the 2MASS $\mj\times\jh$\ CMDs. Cluster\,1 has an age $t=20 - 32$\,Myr,
a reddening $\ebv=2.18\pm0.03$ and a distance to the Sun $\ds=1.64\pm0.19$\,kpc. 
For Cluster\,2 we derive $t=400 - 630$\,Myr, $\ebv=0.91\pm0.03$ and 
$\ds=2.19\pm0.21$\,kpc. For Cluster\,3, $t=32 - 100$\,Myr, $\ebv=0.94\pm0.03$ and 
$\ds=3.93\pm0.35$\,kpc. At the above distances, the core radii for Clusters\,1, 2
and 3 turn out to be $0.32\pm0.03$\,pc, $0.68\pm0.06$\,pc and $0.21\pm0.02$\,pc, 
respectively. The linear limiting diameters are $2.6\pm0.3$\,pc, $4.4\pm0.4$\,pc and 
$3.8\pm0.3$\,pc, respectively, thus suggesting dynamical evolution effects for the 
older clusters. 

For Clusters\,1 and 2, a MF fit $\phi(m)\propto m^{-(1+\chi)}$ resulted in a slope
$\chi=0.85$ and 1.04, respectively. Thus, the observed stellar mass in Cluster\,1 
is $147\,\ms$. In Cluster\,2, including the evolved stars, it is $89\,\ms$. 
The stellar mass in Cluster\,3 is $\sim55\,\ms$. Extrapolating the MF fits down 
to the theoretical low-mass end $m_{low}=0.08\,\ms$, the masses of Clusters\,1 and 
2 turn out to be $382\,\ms$\ and $614\,\ms$, respectively.

The present study suggests that more optical open clusters are yet
to be found in the Galaxy.

\begin{acknowledgements}
This publication makes use of data products from the Two Micron All Sky Survey, which is a joint 
project of the University of Massachusetts and the Infrared Processing and Analysis Center/California 
Institute of Technology, funded by the National Aeronautics and Space Administration and the National 
Science Foundation. We employed catalogues from CDS/Simbad (Strasbourg) and Digitized Sky Survey 
images from the Space Telescope Science Institute (U.S. Government grant NAG W-2166) obtained using 
the extraction tool from CADC (Canada). We also made use of the WEBDA open cluster database. We 
acknowledge support from the Brazilian Institution CNPq.
\end{acknowledgements}

%sssssssssssssssssssssssssssss REFERENCESsssssssssssssssssssssssssssssss
%

\end{document}